\documentclass{tlp}

\jnlPage{1}{?}
\jnlDoiYr{2022}
\doival{10.1017/xxxxx}

\usepackage[utf8]{inputenc}
\usepackage{hyperref}
\usepackage{mathptmx}
\usepackage{amsmath}
\usepackage{amssymb}
\usepackage{stmaryrd} 

\usepackage{listings}
\lstnewenvironment{curry}[1][]{\lstset{#1}}{}
\lstnewenvironment{prolog}{}{}
\newcommand{\listline}{\vrule width0pt depth1.75ex}

\newtheorem{example}{Example} 
\newtheorem{definition}{Definition} 
\newtheorem{theorem}{Theorem} 

\renewcommand{\emptyset}{\varnothing}

\renewcommand{\tt}{\ttfamily}
\newcommand{\codefont}{\small\tt}
\newcommand{\code}[1]{\mbox{\codefont{#1}}} 
\newcommand{\ccode}[1]{``\code{#1}''}  
\newcommand{\us}{\char95} 
\newcommand{\emptygoal}{\mbox{\tiny$\Box$}} 

\def\narr{\leadsto} 
\def\narrstar{\buildrel {*} \over \narr} 
\def\res{\vdash} 
\def\resstar{\vdash^{*}} 
\newcommand{\trans}[2]{\ensuremath{\left\llbracket #1
                       \right\rrbracket _{\cal #2}}}
\newcommand{\transt}[1]{\trans{#1}{T}} 
\newcommand{\transc}[1]{\trans{#1}{C}} 
\newcommand{\transfr}[2]{\trans{#1}{F}^{#2}} 
\newcommand{\transf}[1]{\transfr{#1}{\respos}} 
\newcommand{\transd}[1]{\trans{#1}{D}^{\respos}} 
\newcommand{\respos}{{\cal R}}  
\newcommand{\dt}{{\cal T}}  

\begin{document}

\title{From Logic to Functional Logic Programs}

\lefttitle{Michael Hanus}

\begin{authgrp}
\author{Michael Hanus}
\affiliation{Institut f\"ur Informatik, CAU Kiel, Germany. \\
        \email{mh@informatik.uni-kiel.de}}
\end{authgrp}

\maketitle

\begin{abstract}
Logic programming is a flexible programming paradigm
due to the use of predicates without a fixed data flow.
To extend logic languages with the compact notation
of functional programming, there are various proposals
to map evaluable functions into predicates
in order to stay in the logic programming framework.
Since amalgamated functional logic languages offer
flexible as well as efficient evaluation strategies,
we propose an opposite approach in this paper.
By mapping logic programs into functional logic programs
with a transformation based on inferring
functional dependencies,
we develop a fully automatic transformation
which keeps the flexibility of logic programming
but can improve computations by reducing
infinite search spaces to finite ones.

\medskip
\noindent
\textit{Under consideration for acceptance in TPLP}.
\end{abstract}


\section{Motivation}
\label{sec:motivation}

Functional and logic programming are the most prominent declarative
programming paradigms.
Functional programming provides a compact notation, due to nested expressions,
and demand-driven (optimal) evaluation strategies,
whereas logic programming provides flexibility due to
free variables, unification, and built-in search.
Thus, both paradigms have their advantages for application programming
so that it is reasonable to offer their features in a single language.
One option to achieve this is to
extend a logic language with functional syntax
(equations, nested functional expressions)
and transform evaluable functions into predicates
and flatten nested expressions
\citep{BarbutiBelliaLeviMartelli84,CasasCabezaHermenegildo06,Naish91}.
Hence, logic programming can be considered as the more general paradigm.
On the other hand, functional programming supports efficient, in particular,
optimal evaluation strategies, by exploiting functional dependencies
during evaluation so that it provides more modularity
and new programming concepts, like programming with
infinite data structures \citep{Hughes90}.
In Prolog systems which support coroutining, i.e., delaying
the evaluation of literals when arguments are not sufficiently
instantiated, one can exploit coroutining to implement the
basic idea of lazy evaluation
\citep{CasasCabezaHermenegildo06,Naish91}.
However, the use of coroutines might add new problems.
For instance, computations with coroutining introduce
the risk of incompleteness by floundering \citep{Lloyd87}.
Furthermore, coroutining might yield infinite search spaces
due to delaying literals generated by recursively defined predicates
\citep{Hanus95JLP}.
Although the operational behavior of particular programs
can be improved with coroutines, there are no general results
characterizing classes of programs where this always leads
to an improved operational behavior.

Amalgamated \emph{functional logic languages} \citep{AntoyHanus10CACM}
are an approach to support flexible
as well as efficient evaluation strategies.
For instance, the language Curry \citep{Hanus16Curry}
is based on an optimal and logically sound and complete strategy
\citep{AntoyEchahedHanus00JACM,Antoy97ALP}.
The motivation of this paper is to show that functional logic languages
are actually superior to pure logic languages.
Instead of transforming functions into predicates,
we present a sequence of transformations which map logic programs
into functional logic programs.
Using a sophisticated mapping based on inferring
functional dependencies, we obtain a fully automatic transformation tool
which can reduce the computational efforts.
In particular, infinite search spaces w.r.t.\ logic programs
can be reduced to finite ones w.r.t.\ the transformed
functional logic programs.

This paper is structured as follows.
The next section reviews basic notions of logic and
functional logic programming.
Section~\ref{sec:constrans} presents a simple embedding
of logic into functional logic programs, which is
extended in Sect.~\ref{sec:functrans} and~\ref{sec:demandtrans}
by considering functional dependencies.
Section~\ref{sec:inferresulpos} discusses the inference
of such dependencies and Sect.~\ref{sec:extensions}
adds some extensions. This final transformation is
implemented in a tool sketched in Sect.~\ref{sec:impl}.
We evaluate our transformation in Sect.~\ref{sec:eval}
before we conclude.
The appendix contains the proofs of the theorems.

\section{Logic and Functional Logic Programming}
\label{sec:lp-flp}

In this section we fix our notation for logic programs
and briefly review some relevant features of
functional logic programming.
More details are in the textbook of \cite{Lloyd87} and in surveys on
functional logic programming \citep{AntoyHanus10CACM,Hanus13}.

In logic programming (we use Prolog syntax for concrete examples),
\emph{terms} are constructed from variables ($X,Y\ldots$),
numbers, atom constants ($c,d,\ldots$),
and functors or term constructors ($f,g,\ldots$) applied to
a sequence of terms, like $f(t_1,\ldots,t_n)$.
A \emph{literal} $p(t_1,\ldots,t_n)$ is a predicate $p$ applied
to a sequence of terms, and
a \emph{goal} $L_1,\ldots,L_k$ is a sequence of literals,
where $\emptygoal$ denotes the empty goal ($k = 0$).
Predicates are defined by \emph{clauses} \code{$L$ :- $B$},
where the \emph{head} $L$ is a literal and the \emph{body} $B$ is a goal
(a \emph{fact} is a clause with an empty body $\emptygoal$,
otherwise it is a \emph{rule}).
A \emph{logic program} is a sequence of clauses.

Logic programs are evaluated by SLD-resolution steps,
where we consider the leftmost selection rule here.
Thus, if $G = L_1,\ldots,L_k$ is a goal and
\code{$L$ :- $B$} is a variant of a program clause (with fresh variables)
such that there exists
a most general unifier\footnote{Substitutions, variants, and unifiers are
defined as usual \citep{Lloyd87}.}
(\emph{mgu}) $\sigma$ of $L_1$ and $L$,
then $G \res_\sigma \sigma(B, L_2,\ldots,L_k)$ is
a \emph{resolution} step.
We denote by $G_1 \resstar_\sigma G_m$ a sequence
$G_1 \res_{\sigma_1} G_2 \res_{\sigma_2} \ldots \res_{\sigma_{m-1}} G_m$
of resolution steps with $\sigma = \sigma_{m-1} \circ \dots \circ \sigma_1$.
A \emph{computed answer} for a goal $G$ is a substitution $\sigma$
(usually restricted to the variables occurring in $G$)
with $G \resstar_\sigma \emptygoal$.

\begin{example}\label{ex:concdup-lp}
The following logic program defines the well-known predicate \code{app}
for list concatenation, a predicate \code{app3} to concatenate three lists,
and a predicate \code{dup} which is satisfied if the second argument
occurs at least two times in the list provided as the first argument:
\begin{curry}
app([],Ys,Ys).
app([X|Xs],Ys,[X|Zs]) :- app(Xs,Ys,Zs).$\listline$
app3(Xs,Ys,Zs,Ts) :- app(Xs,Ys,Rs), app(Rs,Zs,Ts).$\listline$
dup(Xs,Z) :- app3(_,[Z|_],[Z|_],Xs).
\end{curry}
The computed answers for the goal \code{dup([1,2,2,1],Z)}
are $\{\code{Z} \mapsto \code{1}\}$ and $\{\code{Z} \mapsto \code{2}\}$.
They can be computed by a Prolog system, but after showing these
answers, Prolog does not terminate due to an infinite search space.
Actually, Prolog does not terminate for the goal \code{dup([],Z)}
since it enumerates arbitrary long lists for the first argument
of \code{app3}.
\end{example}
Functional logic programming \citep{AntoyHanus10CACM,Hanus13}
integrates the most important features
of functional and logic languages in order to provide a variety
of programming concepts.
Functional logic languages
support higher-order functions and lazy (demand-driven) evaluation
from functional programming
as well as non-deterministic search and
computing with partial information from logic programming.
The declarative multi-paradigm language Curry \citep{Hanus16Curry},
which we use in this paper,
is a functional logic language with advanced programming concepts.
Its syntax is close to Haskell \citep{PeytonJones03Haskell},
i.e., variables and names of defined operations
start with lowercase letters and
the names of data constructors start with an uppercase letter.
The application of an operation $f$
to $e$ is denoted by juxtaposition (``$f~e$'').

In addition to Haskell,
Curry allows \emph{free} (\emph{logic}) \emph{variables}
in program rules (equations) and initial expressions.
Function calls with free variables are evaluated by a possibly
non-deterministic instantiation of arguments.

\begin{example}\label{ex:concdup-curry}
The following Curry program\footnote{%
We simplify the concrete syntax by omitting the declaration
of free variables, like \code{z}, which is required in concrete Curry programs
to enable some consistency checks by the compiler.}
defines the operations of Example~\ref{ex:concdup-lp}
in a functional manner, where logic features
(free variables \code{z} and \code{\us})
are exploited to define \code{dup}:
\begin{curry}
app []     ys = ys
app (x:xs) ys = x : app xs ys$\listline$
app3 xs ys zs = app (app xs ys) zs$\listline$
dup xs | xs =:= app3 _ (z:_) (z:_)
       = z
\end{curry}
\ccode{|} introduces a condition, and
\ccode{=:=} denotes semantic unification, i.e.,
the expressions on both sides are evaluated before unifying them.
\end{example}
Since \code{app} can be called with free variables in arguments,
the condition in the definition of \code{dup}
is solved by instantiating \code{z} and
the anonymous free variables \ccode{\us} to appropriate \emph{values}
(i.e., expressions without defined functions)
before reducing the function calls.
This corresponds to narrowing \citep{Reddy85,Slagle74}.
$t \narr_\sigma t'$ is a \emph{narrowing step}
if there is some non-variable position $p$ in $t$,
an equation (program rule) $l \;\code{=}\; r$,
and an mgu $\sigma$ of $t|_p$ and $l$
such that $t' = \sigma(t[r]_p)$,\footnote{We use common notations
from term rewriting \citep{BaaderNipkow98,Terese03}.}
i.e., $t'$ is obtained from $t$ by replacing
the subterm $t|_p$ by the equation's right-hand side and applying
the unifier.
This definition also applies to
conditional equations $l \;\code{|}\; c \;\code{=}\; r$
which are considered as syntactic sugar for the
unconditional equation $l \;\code{=}\; c \;\code{\&>}\; r$,
where the operation \ccode{\&>} is defined by
\code{True$\;$\&>$\;$x = x}.

Curry is based on the
\emph{needed narrowing strategy} \citep{AntoyEchahedHanus00JACM}
which also uses non-most-general unifiers in narrowing steps
to ensure the optimality of computations.
Needed narrowing is a demand-driven evaluation strategy,
i.e., it supports computations with infinite data structures \citep{Hughes90}
and can avoid superfluous computations.
The latter property is our motivation to transform
logic programs into Curry programs, since
this can reduce infinite search spaces to finite ones.
For instance, the evaluation of the expression \code{dup$\;$[]}
has a finite computation space: the generation of longer lists
for the first argument of \code{app3} is avoided since there is
no demand for such lists.

\code{dup} is a \emph{non-deterministic operation}
since it might deliver more than one result for a given argument,
e.g., the evaluation of \code{dup$\,$[1,2,2,1]} yields the values
\code{1} and \code{2}.
Non-deterministic operations, which can formally be
interpreted as mappings from values into sets of values \citep{GonzalezEtAl99},
are an important feature of contemporary functional logic languages.
For the transformation described in this paper,
this feature has the advantage that it is not important
to transform predicates into purely mathematical functions
(having at most one result for a given combination of arguments).

Curry has many more features which are useful for application programming,
like \emph{set functions}  \citep{AntoyHanus09} to encapsulate search,
and standard features from functional programming,
like modules or monadic  I/O \citep{Wadler97}.
However, the kernel of Curry described so far should be sufficient
to understand the transformation described in the following sections.

\section{Conservative Transformation}
\label{sec:constrans}

Functional logic programming is an extension of pure logic programming.
Hence, there is a straightforward way to map logic programs
into functional logic programs:
map each predicate into a Boolean function
and transform each clause into a (conditional) equation.
We call this mapping the \emph{conservative transformation}
since it keeps the basic structure of derivations.
Since narrowing-based functional logic languages
support free variables as well as overlapping rules,
this mapping does not change the set of computed solutions
(in contrast to a purely functional target language where
always the first matching rule is selected).

As a first step to describe this transformation,
we have to map terms from logic into functional logic notation.
Since terms in logic programming (here: Prolog syntax)
have a direct correspondence to data terms (as used in Curry),
the mapping of terms is just a change of syntax
(e.g., uppercase variables are mapped into lowercase,
and lowercase constants and constructors are mapped into
their uppercase equivalents).
We denote this \emph{term transformation} by $\transt{\cdot}$,
i.e., $\transt{\cdot}$ is a syntactic transformation which
maps a (Prolog) term, written inside the brackets,
into a (Curry) data term.
It is defined by a case distinction as follows:
\[
\begin{array}{@{}r@{~~}c@{~~}l@{~~~}l}
\transt{X} & = & x & \mbox{(variable)} \\
\transt{n} & = & n & \mbox{(number constant)} \\
\transt{c} & = & C & \mbox{(atom constant)} \\
\transt{f(t_1,\ldots,t_n)} & = & F~\transt{t_1} ~\ldots~ \transt{t_1}
 & \mbox{(constructed term)}
\end{array}
\]
Based on this term transformation, we define a mapping $\transc{\cdot}$
from logic into functional logic programs where
facts and rules are transformed into
unconditional and conditional equations, respectively
(the symbol \ccode{\&\&} is an infix operator in Curry
denoting the Boolean conjunction):
\[
\begin{array}{@{}r@{~~}c@{~~}l@{~~~}l}
\transc{p(t_1,\ldots,t_n)} & = & p~\transt{t_1} ~\ldots~ \transt{t_n}
 & \mbox{(literal)} \\[1ex]
\transc{l_1,\ldots,l_k} & = &
\transc{l_1} ~\code{\&\&}~\ldots~\code{\&\&}~ \transc{l_k}
 & \mbox{(goal)} \\[1ex]
\transc{l\code{.}} & = & \transc{l} \code{~= True}
 & \mbox{(fact)} \\[1ex]
\transc{l \code{~:-~} b.} & = & \transc{l} \code{~|~} \transc{b} \code{~= True}
 & \mbox{(rule)} \\[1ex]
\transc{clause_1 \,\ldots\, clause_m} & = &
\transc{clause_1} \,\ldots\, \transc{clause_m}
 & \mbox{(program)}
\end{array}
\]
\begin{example}\label{ex:plus-lp}
Consider the following program to add two
natural numbers in Peano representation, where \code{o}
represents zero and \code{s} represents the successor of a natural
\citep{SterlingShapiro94}:
\begin{prolog}
plus(o,Y,Y).
plus(s(X),Y,s(Z)) :- plus(X,Y,Z).
\end{prolog}
The conservative transformation produces the following Curry program:
\begin{curry}
plus O y y = True
plus (S x) y (S z) | plus x y z = True
\end{curry}
Note that the first rule would not be allowed in functional languages,
like Haskell, since the left-hand side is not linear
due to the two occurrences of the pattern variable \code{y}.
For compatibility with logic programming,
such multiple occurrences of variables in patterns are allowed in
Curry, where they are considered as syntactic sugar for
explicit unification constraints.
Thus, the first rule is equivalent to
\begin{curry}
plus O y y' | y =:= y' = True
\end{curry}
\end{example}
Apart from small steps to handle conditions and conjunctions,
there is a strong correspondence between
the derivations steps in the logic programs and the functional logic programs
obtained by the conservative transformation.
Therefore, the following result can be proved by induction
on the length of the resolution and narrowing derivations, respectively.

\begin{theorem}[Correctness of the conservative transformation]
\label{theo-correctness-conservative}
Let $P$ be a logic program and $G$ a goal.
There is a resolution derivation $G \resstar_\sigma \emptygoal$
w.r.t.\ $P$
if and only if
there is a narrowing derivation $\transc{G} \narrstar_{\sigma} \code{True}$
w.r.t.\ $\transc{P}$.
\end{theorem}

\section{Functional Transformation}
\label{sec:functrans}

The conservative transformation simply maps $n$-ary predicates
into $n$-ary Boolean functions.
In order to exploit features from \emph{functional} logic programming,
one should mark at least one argument as a \emph{result argument}
with the intended meaning that the operation maps values for
the remaining arguments into values for the result arguments.
For instance, consider the predicate \code{plus}
defined in Example~\ref{ex:plus-lp}.
Here, the third argument could be considered as a result
argument since \code{plus} maps values for the first two
arguments into a value for the third argument.
Hence, the definition of \code{plus} can also be
transformed into the following functional logic program:
\begin{curry}
plus O y = y
plus (S x) y | z =:= plus x y = S z
\end{curry}
We call this the \emph{functional transformation}
and denote it by $\transfr{\cdot}{}$.
At this point we do not replace the occurrence of \code{z}
in the right-hand side by \code{plus x y} since this
might lead to a different semantics, as we will see later.

It is interesting to note that, without such a replacement,
it is not really relevant which arguments are considered as results.
For instance, if the first two arguments of \code{plus}
are marked as result arguments, we obtain the following
program by the functional transformation:
\begin{curry}
plus y = (O, y)
plus (S z) | (x,y) =:= plus z = (S x, y)
\end{curry}
This specifies a non-deterministic operation which returns,
for a given natural number $n$, all splittings into two numbers
such that their sum is equal to $n$:
\begin{curry}
>  plus (S (S O))
(S (S O), O)
(S O, S O)
(O, S (S O))
\end{curry}
Later we will show how to prefer the transformation into
deterministic operations, since they will lead to
a better operational behavior.
For the moment we should keep in mind that the selection
of result arguments could be arbitrary.
In order to fix it, we assume that, for each $n$-ary predicate $p$,
there is an assignment
$\respos(p/n) \subseteq \{1,\ldots,n\}$
which defines the result argument positions, e.g.,
$\respos(\code{plus}/3) = \{3\}$ or $\respos(\code{plus}/3) = \{1,2\}$.
In practice, one can specify the result argument positions
for a predicate by a specific directive, e.g.,
\begin{prolog}
:- function plus/3: 3.
\end{prolog}
or
\begin{prolog}
:- function plus/3: [1,2].
\end{prolog}
If the set of result argument positions for an $n$-ary predicate is $\{n\}$,
it can also be omitted in the directive, as in
\begin{prolog}
:- function plus/3.
\end{prolog}
Our tool, described below, respects such directives
or tries to infer them automatically.

With these prerequisites in mind, we denote the
\emph{functional transformation w.r.t.\ a result argument position mapping}
$\respos$ by $\transf{\cdot}$.
To define this transformation,
we use the following notation to split the arguments
of a predicate call $p(t_1,\ldots,t_n)$ into the result and the remaining
arguments.
If $\respos(p/n) = \{\pi_1,\ldots,\pi_u\}$ and
$\{\pi'_1,\ldots,\pi'_v\} = \{1,\ldots,n\} \setminus \respos(p/n)$
(where $\pi_i < \pi_{i+1}$ and $\pi'_j < \pi'_{j+1}$),
then the result arguments are
$r_i = t_{\pi_i}$, for $i \in \{1,\ldots,u\}$, and
the remaining arguments are
$a_j = t_{\pi'_j}$, for $j \in \{1,\ldots,v\}$.
Then $\transf{\cdot}$ is defined as follows:
\[
\begin{array}{@{}r@{~}c@{~}l@{~~}l}
\transf{p(t_1,\ldots,t_n)} & = &
\code{(}\transt{r_1},\ldots,\transt{r_u}\code{)~=:=~}
p\,\transt{a_1} \,\ldots\, \transt{a_v}
 & \mbox{(literal, $u>0$)} \\[1ex]
\transf{p(t_1,\ldots,t_n)} & = &
p\,\transt{t_1} \,\ldots\, \transt{t_n}
 & \mbox{(literal, $u=0$)} \\[1ex]

\transf{l_1,\ldots,l_k} & = &
\transf{l_1} ~\code{\&\&}~\ldots~\code{\&\&}~ \transf{l_k}
 & \mbox{(goal)} \\[1ex]

\transf{p(t_1,\ldots,t_n)\code{.}} & = &
p\,\transt{a_1} \,\ldots\, \transt{a_v} \code{~=~}
\code{(}\transt{r_1},\ldots,\transt{r_u}\code{)}
 & \mbox{(fact, $u>0$)} \\[1ex]
\transf{p(t_1,\ldots,t_n)\code{.}} & = &
p\,\transt{t_1} \,\ldots\, \transt{t_n} \code{~=~True}
 & \mbox{(fact, $u=0$)} \\[1ex]

\transf{p(t_1,\ldots,t_n) \code{~:-~} b.} & = &
p\,\transt{a_1} \,\ldots\, \transt{a_v} \code{~|~} \transf{b} \code{~=~}
\code{(}\transt{r_1},\ldots,\transt{r_u}\code{)}
 & \mbox{(rule, $u>0$)} \\[1ex]
\transf{p(t_1,\ldots,t_n) \code{~:-~} b.} & = &
p\,\transt{t_1} \,\ldots\, \transt{t_n} \code{~|~} \transf{b} \code{~=~True}
 & \mbox{(rule, $u=0$)} \\[1ex]

\transf{cls_1 \,\ldots\, cls_m} & = &
\transf{cls_1} \,\ldots\, \transf{cls_m}
 & \mbox{(program)}
\end{array}
\]
As already mentioned, the actual selection of result positions
is not relevant so that we have the following result,
which can be proved similarly to Theorem~\ref{theo-correctness-conservative}:

\begin{theorem}[Correctness of the functional transformation]
\label{theo-correctness-functional}
Let $P$ be a logic program, $\respos$ a result argument position mapping
for all predicates in $P$, and $G$ a goal.
There is a resolution derivation $G \resstar_\sigma \emptygoal$
w.r.t.\ $P$
if and only if
there is a narrowing derivation $\transf{G} \narrstar_{\sigma} \code{True}$
w.r.t.\ $\transf{P}$.
\end{theorem}

\section{Demand Functional Transformation}
\label{sec:demandtrans}

Consider the result of the functional transformation of \code{plus}
w.r.t.\ the result argument position mapping $\respos(\code{plus}/3) = \{3\}$:
\begin{curry}
plus O     y = y
plus (S x) y | z =:= plus x y = S z
\end{curry}
Since the value of \code{z} is determined by the expression
\code{plus$\;$x$\;$y},
we could be tempted to replace the unification in the condition
by a local binding for \code{z}:
\begin{curry}
plus O     y = y
plus (S x) y = S z    where z = plus x y
\end{curry}
Since \code{z} is used only once, we could inline the definition of \code{z}
and obtain the purely functional definition
\begin{curry}
plus O     y = y
plus (S x) y = S (plus x y)
\end{curry}
Although this transformation looks quite natural,
there is a potential problem with this transformation.
In a strict language, where arguments are evaluated before
jumping into the function's body (``call by value''),
there is no difference between these versions of \code{plus}.
However, there is also no operational advantage of this transformation.
An advantage could come from the non-strict or demand-driven
evaluation of functions, as used in Haskell or Curry
and discussed by \cite{Hughes90} and \cite{HuetLevy91}.
For instance, consider the predicate \code{isPos} which returns \code{True}
if the argument is non-zero
\begin{curry}
isPos O     = False
isPos (S x) = True
\end{curry}
and the expression \code{isPos$\;$(plus$\;n_1\;n_2$)},
where $n_1$ is a big natural number.
A strict language requires $n_1 + 1$ rewrite steps to evaluate
this expression, whereas a non-strict language needs only two steps
w.r.t.\ the purely functional definition of \code{plus}.

The potential problem of this transformation comes from the fact
that it does not require the evaluation of subexpressions
which do not contribute to the overall result.
For instance, consider the functions
\begin{prolog}
dec (S x) = x                  const x y = x
\end{prolog}
and the expression $e =\;$\code{const$\;$O$\;$(dec$\;$O)}.
Following the mathematical principle of ``replacing equals by equals'',
$e$ is equivalent to \code{O},
but a strict language does not compute this value.

Hence, it is a matter of taste whether we want to stick
to purely equational reasoning, i.e., ignore the evaluation
of subexpressions that do not contribute to the result,
or strictly evaluate all subexpressions independent of
their demand.\footnote{Note that
every reasonable programming language requires the
non-strict evaluation of conditional (if-then-else) expressions
so that there is no completely strict language.}
Since non-strict evaluation yields reduced search spaces
(as discussed below), we accept this slight change
in the semantics and define the
\emph{demand functional transformation} $\transd{\cdot}$
as follows.
Its definition is identical to $\transf{\cdot}$
except for the translation of a literal in a goal.
Instead of a unification,
$\transd{\cdot}$ generates a local binding
$\code{let/where (}\transt{r_1},\ldots,\transt{r_u} \code{)~=~}
p\,\transt{a_1} \,\ldots\, \transt{a_v}$
if the result arguments $r_1,\ldots,r_u$
are variables which do not occur in the rule's left-hand side
or in result arguments of other goal literals.
The latter restriction avoids inconsistent bindings for a variable.
Since such bindings are evaluated on demand in non-strict languages,
this change has the effect that the transformed programs
might require fewer steps to compute a result.

In order to produce more compact and readable program,
local bindings of single variables, i.e., \code{$x\;$=$\;e$},
are \emph{inlined} if possible, i.e., if there is only
a single occurrence of $x$ in the rule, this occurrence is replaced by $e$
and the binding is deleted.
This kind of inlining is the inverse of the normalisation
of functional programs presented by \cite{Launchbury93}
to specify a natural semantics for lazy evaluation.

\begin{example}\label{ex:app-rev}
Consider  the usual definition of naive reverse:
\begin{prolog}
:- function app/3.
app([],Ys,Ys).
app([X|Xs],Ys,[X|Zs]) :- app(Xs,Ys,Zs).$\listline$
:- function rev/2.
rev([],[]).
rev([X|Xs],Zs) :- rev(Xs,Ys), app(Ys,[X],Zs).
\end{prolog}
The demand functional transformation translates this program
into the Curry program
\begin{curry}
app []     ys = ys
app (x:xs) ys = x : app xs ys$\listline$
rev []     = []
rev (x:xs) = app (rev xs) [x]
\end{curry}
Thanks to the functional \emph{logic} features of Curry,
one can use the transformed program similarly to the logic program.
For instance, the equation \code{app$\;$xs$\;$ys =:= [1,2,3]}
computes all splittings of the list \code{[1,2,3]},
and \code{rev$\;$ps =:= ps} computes palindromes \code{ps}.
\end{example}
An advantage of this transformation becomes apparent
for nested applications of recursive predicates.
For instance, consider the concatenation of three lists,
as shown in Example~\ref{ex:concdup-lp}:\label{def:app3}
\begin{prolog}
:- function app3/4.
app3(Xs,Ys,Zs,Ts) :- app(Xs,Ys,Rs), app(Rs,Zs,Ts).
\end{prolog}
The demand functional transformation maps it into
\begin{curry}
app3 xs ys zs = app (app xs ys) zs
\end{curry}
The Prolog goal \code{app3(Xs,Ys,Zs,[])} has an infinite search space,
i.e., it does not terminate after producing the solution
\code{Xs=[],$\,$Ys=[],$\,$Zs=[]}.
In contrast, Curry has a finite search space since the demand-driven
evaluation avoids the superfluous generation of longer lists.
This shows the advantage of transforming logic programs
into functional logic programs:
the operational behavior is improved, i.e., the size of the search space
could be reduced due to the demand-driven exploration of the search space,
whereas the positive features, like backward computations, are kept.

A slight disadvantage of the demand functional transformation
is the fact that it requires the specification
of the result argument position mapping $\respos$, e.g.,
by explicit \code{function} directives.
In the next section, we show how it can be automatically
inferred.

\section{Inferring Result Argument Positions}
\label{sec:inferresulpos}

As already discussed above, the selection of result arguments
is not relevant for the applicability of our transformation.
For instance, the predicate
\begin{prolog}
p(a,c).
p(b,c).
\end{prolog}
could be transformed into the function
\begin{curry}
p A = C
p B = C
\end{curry}
as well as into the non-deterministic operation
\begin{curry}
p C = A
p C = B
\end{curry}
or just kept as a predicate:
\begin{curry}
p A C = True
p B C = True
\end{curry}
In general, it is difficult to say what is the best representation
of a predicate as a function.
One could argue that deterministic operations are preferable,
but there are also examples where non-deterministic operations
lead to reduced search spaces.
For instance, the complexity of the classical permutation sort
can be improved by defining the computation of permutations
as a non-deterministic operation \citep{GonzalezEtAl99,Hanus13}.

A possible criterion can be derived from the theory
of term rewriting \citep{HuetLevy91} and
functional logic programming \citep{AntoyEchahedHanus00JACM}.
If the function definitions are
\emph{inductively sequential} \citep{Antoy92ALP},
i.e., the left-hand sides of the rules of each function
contain arguments with a unique case distinction,
the demand-driven evaluation (needed narrowing) is optimal
in the number of computed solutions and the length of
successful derivations \citep{AntoyEchahedHanus00JACM}.
In the following, we present a definition of this criterion
adapted to logic programs.

In many logic programs,
there is a single argument which allows a unique case distinction
between all clauses,
e.g., the first argument in the predicates \code{app}, \code{rev},
or \code{plus} shown above.
However, there are also predicates requiring more than one argument
for a unique selection of a matching rule.
For instance, consider Ackermann's function as a logic program,
as presented by \cite{SterlingShapiro94}:\label{ex-ackermann}
\begin{prolog}
ackermann(o,N,s(N)).
ackermann(s(M),o,V) :- ackermann(M,s(o),V).
ackermann(s(M),s(N),V) :- ackermann(s(M),N,V1), ackermann(M,V1,V).
\end{prolog}
The first argument distinguishes between the cases of an atom \code{o}
and a structure \code{s(M)}, but, for the latter case,
two rules might be applicable. Hence, the second argument is necessary
to distinguish between these rules.
Therefore, we call $\{1,2\}$ a set of
\emph{inductively sequential argument positions} for this predicate.

A precise definition of inductively sequential argument positions
is based on the notion of definitional trees \citep{Antoy92ALP}.
The following definition is adapted to our needs.

\begin{definition}[Inductively sequential arguments]
A \emph{partial definitional tree} $\dt$ with a literal $l$ is either
a \emph{clause node} of the form $clause(l ~\code{:-}~ b)$
with some goal $b$, or
a \emph{branch node} of the form $branch(l,p,\dt_1,\ldots,\dt_k)$,
where $p$ is a position of a variable $x$ in $l$,
$f_1,\ldots,f_k$ are pairwise different functors,
$\sigma_i = \{ x \mapsto f_i(x_1,\ldots,x_{a_i}) \}$
where $x_1,\ldots,x_{a_i}$ are new pairwise distinct variables,
and, for all $i$ in $\{1,\ldots,k\}$,
the child $\dt_i$ is a partial definitional tree with literal
$\sigma_i(l)$.

A \emph{definitional tree} of an $n$-ary predicate $p$ defined by a set of
clauses $cs$ is a partial definitional tree $\dt$ with literal
$p(x_1,\ldots,x_n)$, where $x_1,\ldots,x_n$ are pairwise distinct variables,
such that a variant of each clause of $cs$ is represented
in exactly one clause node of $\dt$.
In this case, $p/n$ is called \emph{inductively sequential}.

A set $D \subseteq \{1,\ldots,n\}$ of argument positions of $p/n$ is called
\emph{inductively sequential}
if there is a definitional tree $\dt$ of $p/n$ such that all
positions occurring in branch nodes of $\dt$ with more than one child
are equal or below
a position in $D$.
\end{definition}

The predicate \code{ackermann} shown above has the
inductively sequential argument sets $\{1,2,3\}$ and $\{1,2\}$.
For the predicates \code{app} and \code{rev}
(see Example~\ref{ex:app-rev}),
$\{1\}$ is the minimal set of inductively sequential argument positions.

Our inference of result argument positions
for a $n$-ary predicate $p$ is based on the following heuristic:
\begin{enumerate}
\item
Find a minimal set $D$ of inductively sequential argument positions of $p/n$.
\item
If $D$ exists and the set $R = \{1,\ldots,n\} \setminus D$ is not empty,
select the maximum value $m$ of $R$ as the result argument,
i.e., $\respos(p/n) = \{m\}$, otherwise $\respos(p/n) = \emptyset$.
\end{enumerate}
Thus, a predicate is transformed into a function only if there are
some inductively sequential arguments and some other arguments.
In this case, we select the maximum argument position,
since this is usually the intended one in practical programs.
Moreover, a single result argument allows a better nesting
of function calls which leads to a better demand-driven evaluation.

Minimal sets of inductively sequential argument positions can be computed
by analyzing the heads of all clauses.
There might be different inductively sequential argument sets
for a given set of clauses.
For instance, the predicate \code{q} defined by the clauses
\begin{prolog}
q(a,c).
q(b,d).
\end{prolog}
has two minimal sets of inductively sequential arguments:
$\{1\}$ and $\{2\}$.
The actual choice of arguments is somehow arbitrary.
If predicates are inductively sequential, the results
of \cite{AntoyEchahedHanus00JACM} ensure that the
programs obtained by our transformation with the heuristic described above
can be evaluated in an optimal manner.

A special case of our heuristic to infer result argument positions
are predicates defined by a single rule.
Since such a definition is clearly non-overlapping and inductively sequential,
all such predicates might be considered as functions.
However, this might lead to unintended function definitions, e.g.,
if a predicate is defined by a single clause containing
a conjunction of literals.
Therefore, we use the following heuristic.
A predicate defined by a single rule is transformed
into a function only if the last argument of the head
is not a variable or a variable which occurs in a result argument position
in the rule's body.
The first case is reasonable to transform predicates
which define constants, as
\begin{prolog}
two(s(s(o))).
\end{prolog}
into constant definitions, as
\begin{curry}
two = S (S O)
\end{curry}
An example for the second case is the automatic transformation
of \code{app3} into a function, as shown in Sect.~\ref{def:app3}.

Although these heuristics yield the expected transformations
in most practical cases (they have been developed during the
experimentation with our tool, see below),
one can always override them using an explicit
\code{function} directive in the logic program.

\section{Extensions}
\label{sec:extensions}

Our general objective is the transformation of pure logic programs
into functional logic ones.
Prolog programs often use many impure features which cannot be
directly translated into a language like Curry.
This is intended, because functional logic languages
are an approach to demonstrate how to avoid impure features
and side effects by concepts from functional programming.
For instance, I/O operations, offered in Prolog as
predicates with side effects, can be represented in
functional (logic) languages by monadic operations
which structure effectful computations \citep{Wadler97}.
Encapsulated search (\code{findall}) or cuts in Prolog,
whose behavior depends on the search strategy and ordering of rules,
can be represented in functional logic programming
in a strategy-independent manner as
set functions \citep{AntoyHanus09} or
default rules \citep{AntoyHanus17TPLP}.
Thus, a complete transformation of Prolog programs into Curry programs
might have to distinguish between ``green'' and ``red'' cuts,
which is not computable.
Nevertheless, it is possible to transform some Prolog features
which we discuss in the following.

A useful feature of Prolog is the built-in arithmetic
which avoids to compute with numbers in Peano arithmetic.
For instance, consider the definition of the predicate \code{length}
to relate a list with its number of elements:
\begin{prolog}
length([],0).
length([X|Xs],L) :- length(Xs,L1), L is L1+1.
\end{prolog}
Since the first argument position is inductively sequential,
it is reasonable to transform \code{length} into a function
with $\respos(\code{length}/2) = \{2\}$.
Furthermore, the predicate \ccode{is} evaluates its second argument
to a number and returns this result by unifying it with its
first argument. Thus, $\respos(\code{is}/2) = \{1\}$.
Using this result argument position mapping,
the demand functional transformation yields the program
\begin{curry}
length []     = 0
length (x:xs) = length xs + 1
\end{curry}
where the occurrence of \code{is} is omitted since it behaves
as the identity function.

Arbitrary Prolog cuts cannot be translated (or only into awkward code).
However, Prolog cuts can be avoided by using Prolog's if-then-else construct.
If the condition is a simple predicate, like
a deterministic test or an arithmetic comparison,
it can be translated into a functional if-then-else construct.
For instance, consider the following definition of the
factorial function as a Prolog predicate:
\begin{prolog}
fac(N,F) :- (N=0 -> F=1 ; N1 is N - 1, fac(N1, F1), F is F1 * N).
\end{prolog}
Translating this into a function (note that variable \code{F}
is used in a result argument position in the body)
with the arithmetic operations transformed as discussed above,
we obtain the functional definition
\begin{curry}
fac n = if n == 0 then 1 else fac (n - 1) * n
\end{curry}
With these extensions, many other arithmetic functions
are automatically transformed into their typical functional definition.

\section{Implementation}
\label{sec:impl}

In order to evaluate our approach, we have implemented
the transformation described in this paper
as a tool \code{pl2curry} in Curry so that
it can easily be installed by Curry's package manager.\footnote{%
\url{https://www-ps.informatik.uni-kiel.de/~cpm/pkgs/prolog2curry.html}}
\code{pl2curry} has various options to influence
the transformation strategy (e.g., conservative, without
let bindings, without result position inference, etc).
In the default mode, the demand functional transformation is used
where result arguments are inferred if they are not
explicitly specified by \code{function} directives.
The tool assumes that the logic program is written
in standard Prolog syntax.
It reads the Prolog file and transforms it into an
abstract representation,\footnote{%
\url{https://www-ps.informatik.uni-kiel.de/~cpm/pkgs/prolog.html}}
from which a Curry program is generated
that can be directly loaded into a Curry system.

A delicate practical issue of the transformation
is the typing of the transformed programs.
Curry is a strongly typed language with parametric types
and type classes \citep{WadlerBlott89}.
Since logic programs and standard Prolog do not contain
type information, our tool defines a single type \code{Term}
containing all atoms and functors occurring in the logic program.
Although this works fine for smaller programs,
it could be improved by using type information.
For instance, CIAO-Prolog \citep{HermenegildoEtAl12}
supports the definition of regular and Hindley-Milner types,
which could be translated into algebraic data types,
or \cite{BarbosaFloridoSantosCosta21} describe
a tool to infer similar types from logic programs.
Although there is some interest towards adding types
to Prolog programs \citep{SchrijversSantosCostaWielemakerDemoen08},
there is no general agreement about its syntax and structure.
Therefore, the translation of more refined types
is omitted from the current implementation
but it could be added in the future.

\section{Evaluation}
\label{sec:eval}

The main motivation for this work is to show that
functional logic programs have concrete operational advantages
compared to pure logic programs.
This has been demonstrated by defining transformations
for logic programs into functional logic programs.
The simplest transformations (conservative and functional)
keeps the structure of computations, whereas the
demand functional transformation has the potential advantage
to reduce the computation space by evaluating fewer subexpressions.

The practical comparison of original and transformed programs
is not straightforward since it
depends on the underlying implementation to execute these programs.
Compilers for functional languages
might contain good optimizations since they
must not be prepared for non-deterministic computations
(although Prolog systems based on Warren's Abstract Machine
\citep{Ait-Kaci91,Warren83} implement specific indexing techniques
to support deterministic branching when it is possible).
This can be demonstrated by some typical examples:
the naive reverse of list structures (see Example~\ref{ex:app-rev}),
the highly recursive \code{tak} function used in various
benchmarks \citep{Partain93} for logic and functional languages,
and the Ackermann function (see Sect.~\ref{ex-ackermann}).
Since these logic programs are automatically transformed
into purely functional programs
using our demand functional transformation, we can execute
the original logic programs with Prolog systems and
the transformed programs
with Haskell (GHC) and
Curry (KiCS2 \citep{BrasselHanusPeemoellerReck11}) systems
(since the functional kernel of Curry use the same syntax as Haskell).
KiCS2 compiles Curry programs to Haskell programs
by representing non-deterministic computations as search trees,
i.e., the generated Haskell functions return a tree of all result values.
Table~\ref{table-benchmarks}
contains the average execution times in seconds\footnote{%
The benchmarks, which are contained in the Curry package \code{prolog2curry},
were executed on a Linux machine 
running Debian 10 with an Intel Core i7-7700K (4.2Ghz) processor.
The time is the total run time of executing a binary generated
with the various Prolog/Haskell/Curry systems.}
of reversing a list with 4096 elements,
the function \code{tak} applied to arguments $(27,16,8)$,
implemented with built-in integers (\code{takInt})
and Peano numbers (\code{takPeano}),
and the Ackermann function applied to the Peano representation
of the numbers $(3,9)$.
\begin{table}[ht]
\centering
\caption{Execution times of Prolog, Haskell, and Curry programs}
\label{table-benchmarks}
\begin{tabular}{l@{~~~}rrrr}
\topline
Language: & \multicolumn{1}{c}{Prolog}  & \multicolumn{1}{c}{Prolog}
          & \multicolumn{1}{c}{Haskell} & \multicolumn{1}{c}{Curry} \\
System:   & SWI 8.0.2 & SICStus 4.7.0 & GHC 8.4.4 & KiCS2 3.0.0
\midline
\code{rev{\char95}4096}                         &  0.57~~~ & 0.27~~~ & 0.09~~~ & 0.13~~~ \\
\code{takInt{\char95}27{\char95}16{\char95}8}   &  0.85~~~ & 0.29~~~ & 0.07~~~ & 0.54~~~ \\
\code{takPeano{\char95}27{\char95}16{\char95}8} &  5.81~~~ & 0.79~~~ & 0.17~~~ & 0.47~~~ \\
\code{ackermann{\char95}3{\char95}9}            &231.10~~~ &13.27~~~ & 0.09~~~ & 0.07~~~
\botline
\end{tabular}
\end{table}

Note that the demand strategy has no real advantage in these examples.
The values of all subexpressions are required so that the same
resolution/rewrite steps, possibly in a different order, are performed
in Prolog and Haskell/Curry.
Therefore, the results show the dependency on the
actual language implementations.
Although the table indicates the superiority of the
functional programs (in particular,
GHC seems to implement recursion quite efficiently),
one might also obtain better results for a logic programming
system by sophisticated implementation techniques,
e.g., by specific compilation techniques based on mode information,
as done in Mercury \citep{SomogyiHendersonConway96},
or by statically analyzing programs to optimize the generated code
\citep{VanRoyDespain90}.
For instance, the large execution times of the Prolog version
of the Ackermann function
are probably due to the fact that the function is defined by pattern matching
on two arguments whereas typical Prolog systems implement
indexing on one argument only.
Nevertheless, the results for the Curry system KiCS2 show
a clear improvement without loosing the flexibility of logic programming,
since the same Curry program can compute result values
as well as search for required argument values.
Note that all systems of these benchmarks use unbounded integers,
whereas KiCS2 has a more complex representation of integers
in order to support searching demanded values for free integer variables
\citep{BrasselFischerHuch08}.

Because it is difficult to draw definite conclusions
from the absolute execution times,
we want to emphasize
the qualitative improvement of our transformation.
The demand functional transformation might reduce the number of
evaluation steps and leads to a demand-driven exploration
of the search space.
In the best case, it can reduce infinite search spaces
to finite ones.
We already discussed such examples before.
For instance, our transformation automatically maps
the logic program of Example~\ref{ex:concdup-lp} into the
functional logic program of Example~\ref{ex:concdup-curry}
so that the infinite search space of the predicate \code{dup}
applied to an empty list is cut down to a small finite
search space for the function \code{dup}.
As a similar example, the goal
\code{plus(X,Y,R),$\;$plus(R,Z,o)} (w.r.t.\ Example~\ref{ex:plus-lp})
has an infinite search space, whereas the transformed expression
\code{plus$\;$(plus$\;$x$\;$y)$\;$z =:= O} has a finite search space
w.r.t.\ the demand functional transformation.

These examples show that our transformation has a considerable
advantage when goals containing several recursive predicates
are used to search for solutions.
Such goals naturally occur when complex data structures,
like XML structures, are explored.
The good behavior of functional logic programs on such applications
is exploited by \cite{Hanus11ICLP}
to implement a domain-specific language for XML processing
as a library in Curry.
Without the demand-driven evaluation strategy,
many of the library functions would not terminate.
Actually, the library has similar features as the
logic-based language Xcerpt \citep{BrySchaffert02ICLP}
which uses a specialized unification procedure to ensure
finite matching and unification w.r.t\ XML terms.

\section{Conclusions}
\label{sec:conclusion}

We presented methods to transform logic programs
into functional logic programs.
By specifying one or more arguments as results
to transform predicates into functions
and evaluating them with a demand-driven strategy,
we showed with various examples that this transformation
is able to reduce the computation space.
Although this effect depends on the concrete examples,
our transformation never introduces new or superfluous steps
in successful computations.
We also discussed a heuristic to infer result arguments
for predicates. It is based on detecting
inductively sequential argument positions
so that the programs transformed by our method
benefit from strong completeness and optimality results
of functional logic programming
\citep{AntoyEchahedHanus00JACM,Antoy97ALP}.

In principle, it is not necessary to switch to another
programming language since demand-driven functional computations
can be implemented in Prolog systems supporting coroutining.
However, one has to be careful about the precise evaluation strategy
implemented in this way.
For instance, \cite{Naish91} implements lazy evaluation in Prolog
by representing closures as terms and use \code{when} declarations
to delay insufficiently instantiated function calls.
This might lead to floundering so that
completeness is lost when predicates are transformed into functions.
Moreover, delaying recursively defined predicates could result
in infinite search spaces which can be avoided by complete
strategies \citep{Hanus95JLP}.
\cite{CasasCabezaHermenegildo06} use coroutining to implement
lazy evaluation and offer a notation for functions which are
mapped into Prolog predicates.
Although this syntactic transformation might yield the same values
and search space as functional logic languages, there are no formal results
justifying this transformation.
On the other hand, there are various approaches to implement
lazy narrowing strategies in Prolog
\citep{AntoyHanus00FROCOS,Jimenez-MartinMarino-CarballoMoreno-Navarro92,%
LoogenLopezFraguasRodriguezArtalejo93PLILP,Lopez-FraguasSanchez-Hernandez99}.
In this sense, our results provide a systematic method
to improve computations in logic programs by mapping
predicates into functions and applying sound and complete
evaluation strategies to the transformed programs.
In particular, if predicates are defined with inductively sequential arguments
(as all examples in this paper),
the needed narrowing strategy is optimal,
i.e., the set of computed solutions is minimal
and successful derivations have the shortest possible length
\citep{AntoyEchahedHanus00JACM}.
This does not restrict the flexibility of logic programming
but might reduce the computation space.
Although our implemented tool maps logic programs into Curry programs,
one could also map them back into Prolog programs
by compiling the demand-driven evaluation strategy
into appropriate features of Prolog (e.g., coroutining).

There is a long history to improve the execution
of logic programs by modified control rules
\citep{BruynoogheEtAl89,Narain86}.
However, these proposals usually consider the operational level
so that a declarative justification
(soundness and completeness) is missing.
In this sense, our work provides a justification for specific
control rules used in logic programming,
since it is based on soundness and completeness results
for functional logic programs.

For future work it is interesting
to use a refined representation of types
(as discussed in Sect.~\ref{sec:impl})
or to consider other methods to infer result positions,
e.g., by a program analysis taking into account the data flow
between arguments of literals in goals.


\newpage
\appendix

\section{Proofs}

\emph{Theorem~\ref{theo-correctness-conservative}
(Correctness of the conservative transformation)}\\
Let $P$ be a logic program and $G$ a goal.
There is a resolution derivation $G \resstar_\sigma \emptygoal$
w.r.t.\ $P$
if and only if
there is a narrowing derivation $\transc{G} \narrstar_{\sigma} \code{True}$
w.r.t.\ $\transc{P}$.

\begin{proof}
We generalize the theorem and prove it also for empty goals
$G = \emptygoal$ which are transformed as
\[
\begin{array}{@{}r@{~~}c@{~~}l@{~~~}l}
\transc{\emptygoal} & = & \code{True}
\end{array}
\]
The operations for Boolean conjunction and conditional expressions
are defined in Curry by
\begin{curry}
True  && b = b           True &> x = x
False && b = False
\end{curry}
First, we prove the existence of a narrowing derivation
$\transc{G} \narrstar_{\sigma} \code{True}$ w.r.t.\ $\transc{P}$
for each resolution derivation $G \resstar_\sigma \emptygoal$
w.r.t.\ $P$ by induction on the length $n$ of the given resolution derivation.

Base case ($n = 0$):
Since the derivation has zero length,
$G = \emptygoal$ and $\sigma = \{\}$.
Then $\transc{G} = \code{True}$ and $\code{True} \narrstar_{\{\}} \code{True}$
so that the claim holds in this case.

Inductive case ($n > 0$):
Consider a resolution derivation $G \resstar_\sigma \emptygoal$
w.r.t.\ $P$ consisting of $n>0$ resolution steps.
Hence, this derivation has the structure
\[
G \res_{\sigma'} G' \resstar_{\sigma''} \emptygoal
\]
where $G = l_1,\ldots,l_k$ and
there exists a variant \ccode{$l$ :- $b$.} of a program rule
(the case for a fact \ccode{$l$.} can be proved in a similar way)
such that $\sigma'$ is an mgu of $l$ and $l_1$ and
$G' = \sigma'(b,l_2\,\ldots,l_k)$.
Since $G' \resstar_{\sigma''} \emptygoal$
is a derivation with $n-1$ resolution steps,
there are derivations
\[
\begin{array}{rcl}
\sigma'(b) & \resstar_{\sigma_1} & \emptygoal \\
\sigma_1(\sigma'(l_2)) & \resstar_{\sigma_2} & \emptygoal \\
\vdots \\
\sigma_{k-1}(\cdots(\sigma_1(\sigma'(l_k)))\cdots) & \resstar_{\sigma_k} & \emptygoal
\end{array}
\]
where each derivation has less than $n$ steps and
$\sigma'' = \sigma_k \circ \cdots \circ \sigma_1$.
Hence, by induction hypothesis, there are narrowing derivations
\[
\begin{array}{rcl}
\transc{\sigma'(b)} & \narrstar_{\sigma_1} & \code{True} \\
\transc{\sigma_1(\sigma'(l_2))} & \narrstar_{\sigma_2} & \code{True} \\
\vdots \\
\transc{\sigma_{k-1}(\cdots(\sigma_1(\sigma'(l_k)))\cdots)}
 & \narrstar_{\sigma_k} & \code{True}
\end{array}
\]
w.r.t.\ $\transc{P}$.
By definition of $\transc{\cdot}$,
\[
\transc{G} ~=~ \transc{l_1} ~\code{\&\&}~\ldots~\code{\&\&}~ \transc{l_k}
\]
and
\[
\transc{l} \code{~|~} \transc{b} \code{~= True}
\]
is a variant of a program rule of $\transc{P}$.
Since the conservative transformation changes only the syntax
of literals but keep their structure and variables,
$\sigma'$ is an mgu of $\transc{l_1}$ and $\transc{l}$ so that
\[
\transc{G} \narr_{\sigma'}
(\sigma'(\transc{b} \code{\&>~True})
~\code{\&\&}~ \transc{l_2} ~\code{\&\&}~\ldots~\code{\&\&}~ \transc{l_k}
\]
is a narrowing step.
Due to the existence of the narrowing derivations shown above
and the definition of \code{\&>} and \code{\&\&},
we can combine all this into a narrowing derivation
\[
\transc{G} \narrstar_{\sigma'' \circ \sigma'} \code{True}
\]
Thus, the claim holds.

To prove the opposite direction, we assume the existence of
a narrowing derivation
$\transc{G} \narrstar_{\sigma} \code{True}$ w.r.t.\ $\transc{P}$
and have to show the existence of a
resolution derivation $G \resstar_\sigma \emptygoal$ w.r.t.\ $P$.
We prove this claim by induction on the length $n$
of the given narrowing derivation.

Base case ($n = 0$):
Since the derivation $\transc{G} \narrstar_{\sigma} \code{True}$
has zero length,
$\transc{G} = \code{True}$ and $\sigma = \{\}$ so that $G = \emptygoal$.
Hence, the claim vacuously holds.

Inductive case ($n > 0$):
Assume there is a narrowing derivation $\transc{G} \narrstar_\sigma \code{True}$
w.r.t.\ $\transc{P}$ consisting of $n>0$ narrowing steps.
Hence, this derivation has the structure
\[
\transc{G} \narr_{\sigma'} G' \narrstar_{\sigma''} \code{True}
\]
where the derivation $G' \narrstar_{\sigma''} \code{True}$ consists
$n-1$ narrowing steps.
Therefore $\transc{G} \neq \code{True}$
(since there is no rule in $\transc{P}$ with left-hand side \code{True})
so that $G \neq \emptygoal$.
Thus, $G = l_1,\ldots,l_k$ ($k>0$) and
\[
\transc{G} ~=~ \transc{l_1} ~\code{\&\&}~\ldots~\code{\&\&}~ \transc{l_k}
\]
Furthermore, there is a narrowing step for
$\transc{l_1}$ w.r.t. $\transc{P}$.
By definition of $\transc{P}$,
the logic program contains a variant of
a fact \ccode{$l$.} or a rule \ccode{$l$ :- $b$.}
so that $b' = \code{True}$ or $b' = \code{\transc{b} \&> True}$,
respectively,
such that $\sigma'$ is an mgu of $\transc{l}$ and $\transc{l_1}$ and
\[
G' ~=~ \sigma'(b' ~\code{\&\&}~ \transc{l_2} ~\code{\&\&}~\ldots~\code{\&\&}~ \transc{l_k})
   ~=~ \sigma'(b') ~\code{\&\&}~ \transc{\sigma'(l_2)} ~\code{\&\&}~\ldots~\code{\&\&}~ \transc{\sigma'(l_k)}
\]
By definition of \code{\&\&} and the fact that
$G' \narrstar_{\sigma''} \code{True}$ consists
$n-1$ narrowing steps,
there are narrowing derivations
\[
\begin{array}{rcl}
\transc{\sigma'(b')} & \narrstar_{\sigma_1} & \code{True} \\
\transc{\sigma_1(\sigma'(l_2))} & \narrstar_{\sigma_2} & \code{True} \\
\vdots \\
\transc{\sigma_{k-1}(\cdots(\sigma_1(\sigma'(l_k)))\cdots)}
 & \narrstar_{\sigma_k} & \code{True}
\end{array}
\]
where each derivation has less than $n$ steps and
$\sigma'' = \sigma_k \circ \cdots \circ \sigma_1$.
Hence, by induction hypothesis, there are resolution derivations
\[
\begin{array}{rcl}
\sigma'(b'') & \resstar_{\sigma_1} & \emptygoal \\
\sigma_1(\sigma'(l_2)) & \resstar_{\sigma_2} & \emptygoal \\
\vdots \\
\sigma_{k-1}(\cdots(\sigma_1(\sigma'(l_k)))\cdots) & \resstar_{\sigma_k} & \emptygoal
\end{array}
\]
where $b'' = \emptygoal$ if $b' = \code{True}$ or
$b'' = b$ if $b' = \code{\transc{b} \&> True}$.
Hence,
\[
\sigma'(b'',l_2,\ldots,l_k) ~\resstar_{\sigma_k \circ \cdots \circ \sigma_1}~
\emptygoal
\]
is a resolution derivation.
Furthermore, $G \res_{\sigma'} \sigma'(b'',l_2,\ldots,l_k)$
is a resolution step w.r.t.\ $P$
since $\sigma'$ is also an mgu of $l_1$ and $l$.
Altogether, the resolution derivation
\[
G \resstar_{\sigma'' \circ \sigma'} \emptygoal
\]
exists, which proves the claim.
\end{proof}

\noindent
\emph{Theorem~\ref{theo-correctness-functional}
(Correctness of the functional transformation)}\\
Let $P$ be a logic program, $\respos$ a result argument position mapping
for all predicates in $P$, and $G$ a goal.
There is a resolution derivation $G \resstar_\sigma \emptygoal$
w.r.t.\ $P$
if and only if
there is a narrowing derivation $\transf{G} \narrstar_{\sigma} \code{True}$
w.r.t.\ $\transf{P}$.

\begin{proof}
Note that the difference between the conservative transformation
$\transc{\cdot}$ and the functional transformation $\transf{\cdot}$
is the transformation of literals.
The conservative transformation maps a literal into a structurally
equivalent call of a Boolean function,
whereas the functional transformation maps
literals with a non-empty set of result argument positions into a unification
between the call with some arguments and the tuple of the result
arguments. Since this tuple does not contain any evaluable function,
the unification is equivalent to replacing the function by its result
and unifying this with the given tuple.
Hence, one can prove this theorem similarly to
Theorem~\ref{theo-correctness-conservative}.
\end{proof}

\end{document}